\documentclass[lettersize,journal]{IEEEtran}
\usepackage{empheq}
\usepackage{amsmath,amsfonts}
\usepackage{array}
\usepackage[caption=false,font=normalsize,labelfont=sf,textfont=sf]{subfig}
\usepackage{textcomp}
\usepackage{stfloats}
\usepackage{url}
\usepackage{verbatim}
\usepackage{graphicx}
\usepackage{enumerate}
\usepackage{gensymb}
\usepackage{bm}
\usepackage{CJK}
\usepackage{indentfirst}
\usepackage{cases}
\usepackage{multirow}
\usepackage{booktabs}
\usepackage{cite}
\makeatletter
\newif\if@restonecol
\makeatother

\usepackage[linesnumbered,ruled,lined]{algorithm2e}

\def\BibTeX{{\rm B\kern-.05em{\sc i\kern-.025em b}\kern-.08em
    T\kern-.1667em\lower.7ex\hbox{E}\kern-.125emX}}
\usepackage{balance}
\begin{document}
\title{Highly Accelerated MRI via Implicit Neural Representation Guided Posterior Sampling of Diffusion Models}
\author{Jiayue Chu, Chenhe Du, Xiyue Lin, Yuyao Zhang,~\IEEEmembership{Member,~IEEE,} and Hongjiang Wei,~\IEEEmembership{Member,~IEEE}
\thanks{
(Corresponding author: Hongjiang Wei)

J. Chu, and H. Wei are with the School of Biomedical Engineering, Shanghai Jiao Tong University, Shanghai 200030, China (e-mail: jiayue-chu@sjtu.edu.cn; hongjiang.wei@sjtu.edu.cn).

C. Du, X. Lin and Y. Zhang are with the School of Information Scienceand Technology, ShanghaiTech University, Shanghai 201210, China (e-mail: duchenhe@shanghaitech.edu.cn; linxy@shanghaitech.edu.cn; zhangyy8@shanghaitech.edu.cn).

}}


\markboth{Journal of \LaTeX\ Class Files,~Vol.~18, No.~9, September~2020}%
{Shell \MakeLowercase{\textit{et al.}}: Highly Accelerated MRI via Implicit Neural Representation Guided Posterior Sampling of Diffusion Models}
\maketitle

\begin{abstract}
Reconstructing high-fidelity magnetic resonance (MR) images from under-sampled k-space is a commonly used strategy to reduce scan time.
The posterior sampling of diffusion models based on the real measurement data holds significant promise of improved reconstruction accuracy.
However, traditional posterior sampling methods often lack effective data consistency guidance, leading to inaccurate and unstable reconstructions.
Implicit neural representation (INR) has emerged as a powerful paradigm for solving inverse problems by modeling a signal's attributes as a continuous function of spatial coordinates.
In this study, we present a novel posterior sampler for diffusion models using INR, named DiffINR.
The INR-based component incorporates both the diffusion prior distribution and the MRI physical model to ensure high data fidelity.
DiffINR demonstrates superior performance on experimental datasets with remarkable accuracy, even under high acceleration factors (up to R=12 in single-channel reconstruction).
Notably, our proposed framework can be a generalizable framework to solve inverse problems in other medical imaging tasks.
\end{abstract}


\section{Introduction}
\IEEEPARstart{M}{agnetic} Resonance Imaging (MRI) is a widely used imaging technique, renowned for its exceptional soft tissue contrast, yet hampered by the inherently long scan time.
A common strategy to accelerate MRI scans involves under-sampling k-space data while maintaining the diagnostic quality of the reconstructed images.
This approach is challenging due to the ill-posed nature of sparsely sampled data, making artifact-free reconstruction difficult.
Extensive research has explored various methods to address this challenge, including traditional optimization and deep learning (DL)-based approaches \cite{ref1, ref2}.
Traditional methods, such as compressed sensing (CS), leverage prior knowledge about the inherent sparsity in a transformational domain \cite{ref1,ref3,ref4} or rely on low-rankness \cite{ref5,ref6,ref7} to reconstruct images.
However, these methods face limitations when it comes to achieving higher acceleration rates.
Such limitations stem from the assumptions regarding the prior knowledge that is typically either manually constructed or confined to learned sparse codes, limiting their efficacy and adaptability.

Over the past few years, supervised DL has been regarded as a powerful data-driven approach for solving inverse problems of MRI reconstruction \cite{ref2,ref8,ref9,ref10,ref11}, demonstrating advanced performance in reconstructing highly under-sampled MRI data.
However, these methods need large quantities of high-quality data pairs for training, which is a challenge to obtain in the medical imaging field.
Moreover, the effectiveness of these models tends to decline when applied to out-of-distribution (OOD) conditions, such as different acceleration rates, sampling patterns, and anatomical structures.
To overcome these challenges, data-specific unsupervised DL methods have been proposed, such as deep image prior (DIP).
The DIP framework proposed in \cite{ref12} shows that convolutional neural networks (CNNs) have the intrinsic ability to regularize a variety of ill-posed inverse problems without the need for paired training data.
However, DIP-based methods often suffer from spectral bias \cite{ref13,ref14}, leading to the loss of high-frequency details in reconstructed images.
To address this limitation, coordinate-based implicit neural representation (INR) has emerged as an effective solution \cite{ref15}.
Technically, INR represents the target object as a continuous function of the spatial coordinates, which is parameterized with a fully connected neural network.
Before being fed into the network, the coordinates are transformed through unique encoding techniques into a more resilient and adaptable representation domain (e.g., Fourier domain \cite{ref16}) to reduce spectral bias \cite{ref17,ref18}.
A well-trained INR acts as a continuous function to map image spatial coordinates to corresponding image intensities.
INR offers several notable advantages: (1) a flexible framework for incorporating specific priors or regularizations, and (2) the ability to capture high-frequency details of the representing object \cite{ref16,ref19}.
Previous studies have shown that INR can outperform traditional methods when combined with specific imaging physical priors \cite{ref20,ref21}.
However, similar to the traditional CS-based methods, INR faces a limitation in high acceleration rates due to its reliance on hand-crafted priors, such as total variation (TV) \cite{ref22}.
These hand-crafted priors often struggle to represent complex underlying data distributions accurately.

Recently, the diffusion model has shown its superior capability in MRI acceleration \cite{ref23,ref24}.
Specifically, an unconditional pre-trained diffusion model serves as a generative prior and contains rich knowledge of the complicated MR image structures independent of specific scanning parameters or tissue types.
By leveraging these unconditional diffusion priors, reconstructing specific MR images with under-sampled data to ensure data consistency (DC) can be conceptualized as posterior sampling for diffusion models.
Currently, there are two main categories of posterior sampling methods for DC operations: (1) projecting the generated prior onto the measurement space at each reverse sampling step \cite{ref25,ref26,ref27,ref28} and (2) calculating the posterior probability from the measurements based on certain assumptions \cite{ref29,ref30}.
Although these two types of posterior sampling methods have demonstrated commendable efficacy in MRI reconstruction, they encounter challenges with uncertainty and insufficient DC [31], resulting in inaccurate and instable MRI reconstruction.
This is especially problematic when dealing with highly accelerated rates or OOD data, where these posterior sampling methods are prone to producing erroneous and unstable results (e.g., ring artifacts [27]).
In clinical settings, any inaccuracies and instabilities in image reconstructions could lead to misjudgments by healthcare professionals.

In this study, we develop an INR-based posterior sampling technique for high-acceleration MRI reconstruction called DiffINR.
Inspired by the work in \cite{ref27,ref30}, we introduce INR into the posterior sampling process of diffusion models for DC operation.
Our INR-based posterior sampler maintains DC throughout the diffusion sampling process by leveraging INR's unique ability to integrate diffusion priors with the MRI physical models.
Specifically, we use a well-trained diffusion model to generate prior information about high-quality MRI distribution.
We then employ INR to guarantee the reconstruction DC for specific measurements.
Results show that the INR-based posterior sampler can generate more accurate and stable reconstruction results.
Our contributions are as follows:
\begin{enumerate}[1)]
\item To the best of our knowledge, DiffINR is the first method that explores the potential of combining the diffusion model with INR-based DC operations for solving the inverse problem in MRI reconstruction.
\item Our INR-based module is training-free and can be easily integrated into other diffusion models. Additionally, the INR-based module can be easily combined with specific physical models (e.g., MRI forward physical model) to achieve accurate and stable DC.
\item The quantitative and qualitative results demonstrate that DiffINR effectively improves the accuracy of MRI reconstruction and offers performance comparable to supervised DL methods. 
\end{enumerate}

\section{Related Work}
\subsection{Problem Formulation}
In accelerated MRI acquisition, we formulate the process as the following forward measurement model:
\begin{equation}
\label{deqn_ex1}
y=Ax+e,
\end{equation}
where $x\in\mathbb{C}^{n}$ is the desired MR image, $y\in\mathbb{C}^{m}$ is the measurement data in k-space, $e\in\mathbb{C}^{m}$ represents noise, and $A\in\mathbb{C}^{m\times n}$ denotes the forward acquisition operator.
Specifically, the forward operator can be written as
\begin{equation}
\label{deqn_ex2}
A=MFS,
\end{equation}
where $M$ is the diagonalized sampling mask that represents the sub-sampling operator, $F$ is the Fourier transform matrix and ${S=[S_{1},S_{2},\ldots,S_j]}$ denotes the diagonalized sensitivity map matrix of $j$ different coils.

Recovering $x$ from measurement $y$ can be treated as a posterior sampling process based on the conditional probability density distribution $p(x|y)$.
Thus, the reconstruction problem can be formulated as a maximum posterior probability (MAP) estimation:
\begin{equation}
\label{deqn_ex3}
x_{0}=\underset{x}{\mathrm{argmax}}\log p(x|y)=\underset{x}{\mathrm{argmin}}-\log p(y|x)-\log p(x)
\end{equation}
where $x_{0}$ is the final reconstruction.

To solve the posterior sampling problem by diffusion model, iterative denoising followed by DC operations from timestep $T$ to 0 is employed to get the clean image $x_{0}$ for measurement data $y$.

\subsection{Diffusion Model in MRI Reconstruction}
In recent years, diffusion models have shown great potential in solving the MAP estimation of MRI reconstruction for better accuracy and generalization capability \cite{ref32,ref33,ref34,ref35}.
Jalal \textit{et al.} \cite{ref28} first proposed to use of annealed Langevin Dynamics posterior sampling for multi-channel acquired MRI reconstruction.
Chung \textit{et al.} \cite{ref27} and Song \textit{et al.} \cite{ref25} used score-based diffusion with iterative DC projections, achieving high-quality posterior sampling results.
Based on the idea of DC projection in the sampling phase, Peng \textit{et al.} \cite{ref26} proposed a coarse-to-fine sampling strategy for accelerating MRI reconstruction.
Güngör \textit{et al.} \cite{ref36} introduced the prior adaption operation in diffusion models for fast and reliable reconstructions.

To further improve the reconstruction performance, Cao \textit{et al.} \cite{ref37} proposed a high-frequency space diffusion model to improve the reconstruction stability.
Cui \textit{et al.} \cite{ref38} and Cao \textit{et al.} \cite{ref39} proposed a new paradigm by introducing the iterative self-consistent parallel imaging reconstruction (SPIRiT) \cite{ref40} to the posterior sampling process.
In the k-space domain, Tu \textit{et al.} \cite{ref41} proposed a weighted k-space model for a robust and flexible reconstruction scheme.
Peng \textit{et al.} \cite{ref42} employed hankel-k-space in the generative model to achieve high-quality reconstruction with k-space data of one single subject.

\subsection{Implicit Neural Representation}
The concept of INR first emerged in the context of view synthesis in computer vision.
Mildenhall \textit{et al.} \cite{ref17} pioneered the utilization of a fully connected network to encode a 3D scene as a neural radiance field (NeRF).
Researchers have explored the fusion of INR with various coordinate encoding functions, enabling the capturing of high-frequency image details, such as the Fourier feature map \cite{ref16}, radial encoding \cite{ref43}, and hash encoding \cite{ref19}.
Leveraging multilayer perceptron (MLP) and an appropriate encoding function that maps input coordinates to a high-dimensional space, INR has showcased remarkable efficacy across diverse computer vision tasks \cite{ref44,ref45,ref46}.

Previous work demonstrates INR’s notable capability in addressing inverse problems in medical image reconstruction \cite{ref21,ref47,ref48,ref49}.
For MRI acceleration, Shen \textit{et al.} \cite{ref20} proposed INR learning with prior embedding (NeRP) to get a high-quality result, while it is limited to the case where long-term longitudinal scanning is performed.
Similarly, Feng \textit{et al.} \cite{ref21} proposed an INR approach with sensitivity map estimation (IMJENSE) for parallel MRI.
Although IMJENSE demonstrated relatively high acceleration rates compared to traditional methods, it faces limitations when dealing with highly under-sampled data due to a lack of comprehensive prior knowledge of the complex data distributions.
\begin{figure*}[t!]
  \centering
  \includegraphics[width=1\textwidth]{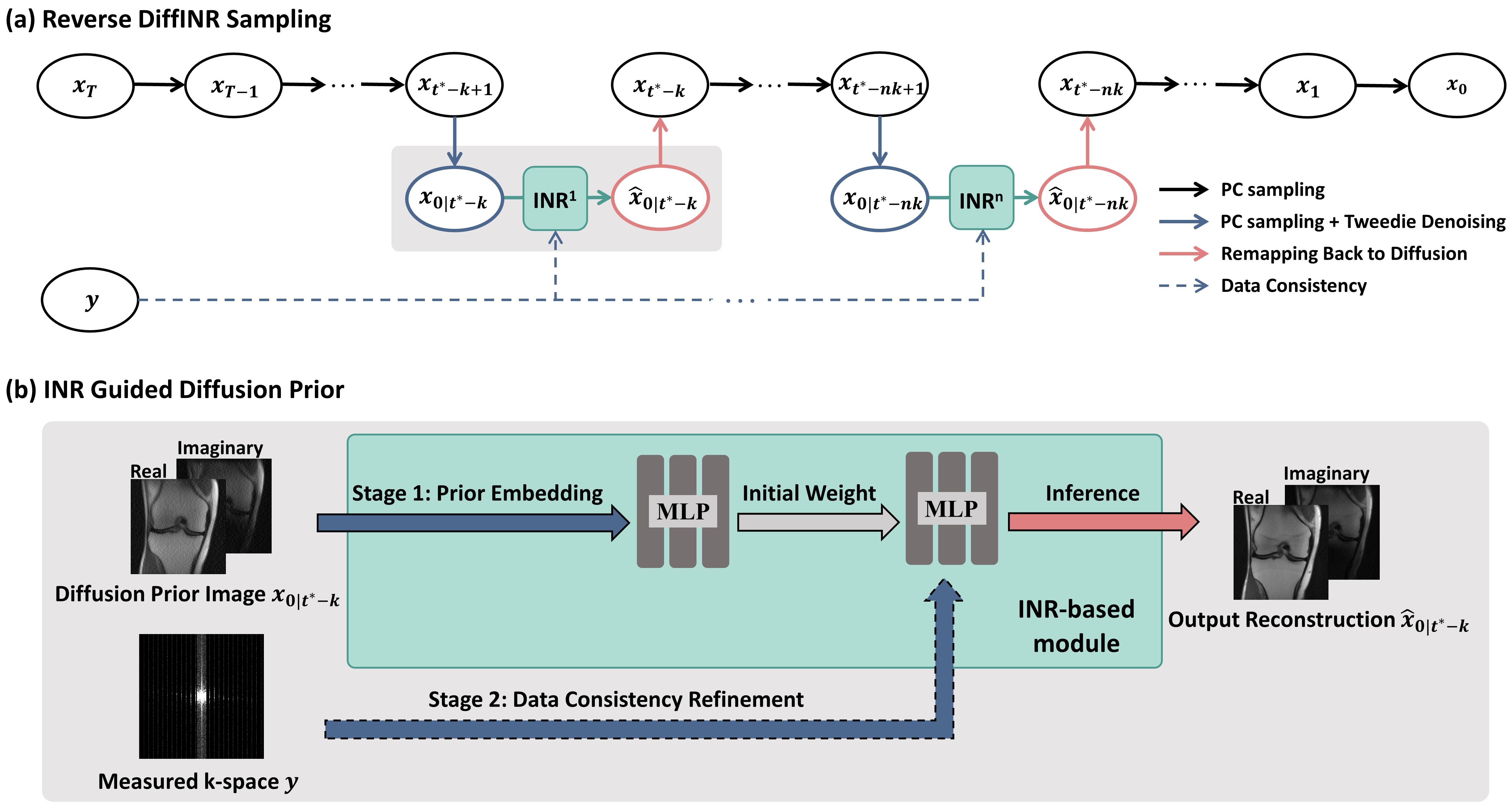}
  \caption{Overview of the proposed DiffINR framework. (a) The reverse sampling process of DiffINR with iterative INR-based modules. The total number of reverse sampling steps is $T$. $t^*$ indicates the INR start timestep and $k$ denotes the INR interval. (b) INR-based posterior sampling for temporal diffusion prior. This process is divided into two stages: In Stage 1, the diffusion prior image is embedded into the INR as an initial weight for following Stage 2. In Stage 2, the INR combines the physical model with the acquired data, guaranteeing data fidelity.}
  \label{fig:XXX}
\end{figure*}

\section{Methodology}
\subsection{Posterior Sampling of Diffusion Models}
For a temporal diffusion sampling at timestep $t$, Eq. (3) can be written as:
\begin{equation}
\label{deqn_ex4}
x_{t-1}=\underset{x_{t}}{\mathrm{argmin}}-\log p_{t}(y|x_{t})-\log p_{t}(x_{t})
\end{equation}
In (4), the unconditional prior distribution $p_{t}(x_{t})$ can be calculated by the pre-trained diffusion model.
Although the measurement model $p_{t}(y|x_{t})$ is theoretically difficult to present in the analytical formulation, the forward physical model from $x_{0}$ to $y$ can be explicitly represented as (1).
To exploit the measurement model $p_{t}(y|x_{0})$, we factorize $p_{t}(y|x_{t})$ as follows:
\begin{equation}
\label{deqn_ex5}
p_{t}(y|x_{t})=\int p_{t}(y|x_{0})p_{t}(x_{0}|x_{t})dx_{0}
\end{equation}
where $p_{t}(x_{0}|x_{t})$ estimates a clean image $x_{0}$ based on noisy image $x_{t}$.
Notably, our method has no constraints on the stochastic differential equation (SDE) in diffusion models, such as variance preserving (VP) SDE or variance exploding (VE) SDE.
Here, we take the VP-SDE as an example.
The predicted $x_{0}$ based on the $x_{t}$ can be written as:
\begin{equation}
\label{deqn_ex6}
x_{D}(x_{t})=x_{0|t}\approx(x_{t}-\sqrt{1-\bar{\alpha}_{t}}s_{\theta}(x_{t},t))/\sqrt{\bar{\alpha}_{t}},
\end{equation}
where $x_{D}(\cdot)$ denotes the denoising process to predict $x_{0|t}$, ${\alpha}_{t}$ is a pre-defined hyperparameter in the pre-trained diffusion model, $s_{\theta}(x_{t},t)$ denotes the predicted noise level of timestep $t$ and $\theta$ is the well-trained model’s parameters.
The denoising process of (6) is called Tweedie denoising and the formula is defined as the Tweedie formula in previous works \cite{ref50,ref51}. 
Based on (5) and (6), we can get:
\begin{equation}
\label{deqn_ex7}
p_{t}(y|x_{t})\approx p_{t}(y|x_{0|t}),
\end{equation}
where $x_{0|t}:=\mathbb{E}[x_{0}|x_{t}]=\mathbb{E}_{x_{0}\sim p(x_{0}|x_{t})}[x_{0}]$.
We can rewrite the posterior sampling problem as
\begin{equation}
  \label{deqn_ex8}
x_{t-1}\approx\underset{x_{t}}{\mathrm{argmin}}-\log p_{t}(y|x_{0|t})-\log p_{t}(x_{t}).
\end{equation}

Assuming the measurement noise $e$ is i.i.d. zero mean, normal distributed and uncorrelated additive, the $p_{t}(y|x_{0|t})$ becomes
\begin{equation}
\label{deqn_ex9}
p_{t}( y|x_{0|t} )=\frac{1}{\sqrt{(2\pi)^{m}\sigma^{2m}}}\exp \left[-\frac{\|y-A(x_{0|t})\|_{2}^{2}}{2\sigma^{2}}\right]
\end{equation}
where $m$ is the dimension of measurement data $y$. the optimization problem (8) can be expressed as:
\begin{equation}
  \label{deqn_ex10}
x_{t-1}=\underset{x_{t}}{\mathrm{argmin}}\left[\parallel y-A(x_{D}(x_{t}))\parallel_{2}^{2}+\lambda H(x_{t})\right],
\end{equation}
where $\lambda=2\sigma^{2}$ is the penalty weight between the DC term and the diffusion generation term.
$H(\cdot)$ presents the unconditional pre-trained diffusion model.

\subsection{Diffusion Prior Generation}
Based on the quadratic penalty approach \cite{ref52}, the problem in (10) can be split into two subproblems: (1) unconditional diffusion prior generation based on the pre-trained $H(\cdot)$; (2) DC operation based on the MRI physical model.
With a well-trained diffusion model, we can obtain the MRI prior distribution $H(x_{t})$ in (10), which is formulated as: 
\begin{equation}
\label{deqn_ex11}
x_{t-1}=\underset{x_{t}}{\mathrm{argmin}}H(x_{t})=(1+\frac{1}{2}\beta_{t})x_{t}+\beta_{t}s_{\theta}(x_{t},t)+\sqrt{\beta_{t}}\epsilon
\end{equation}
where $x_{t-1}$ denotes the unconditional prior diffusion image, $\epsilon\in N(0,I)$ and $\beta_{t}$ is the pre-defined parameters. 

To build the relationship between the diffusion generative prior and the measurement model, we utilize the Tweedie formula in (6) to generate a prior image with reduced noise $x_{0|t-1}=x_{D}(x_{t-1})$.
With $x_{0|t-1}$, the DC operation can be formulated as
\begin{equation}
  \label{deqn_ex12}
\hat{x}_{0|t-1}=\underset{x}{\mathrm{argmin}}\|y-Ax\|_{2}^{2}+\rho\|x_{0|t-1}-x\|_{2}^{2}
\end{equation}
where the first term is used for DC, $\rho$ is a penalty factor and $\hat{x}_{0|t-1}$ denotes the result after DC refinement.

\begin{figure*}[t!]
  \centering
  \includegraphics[width=1\textwidth]{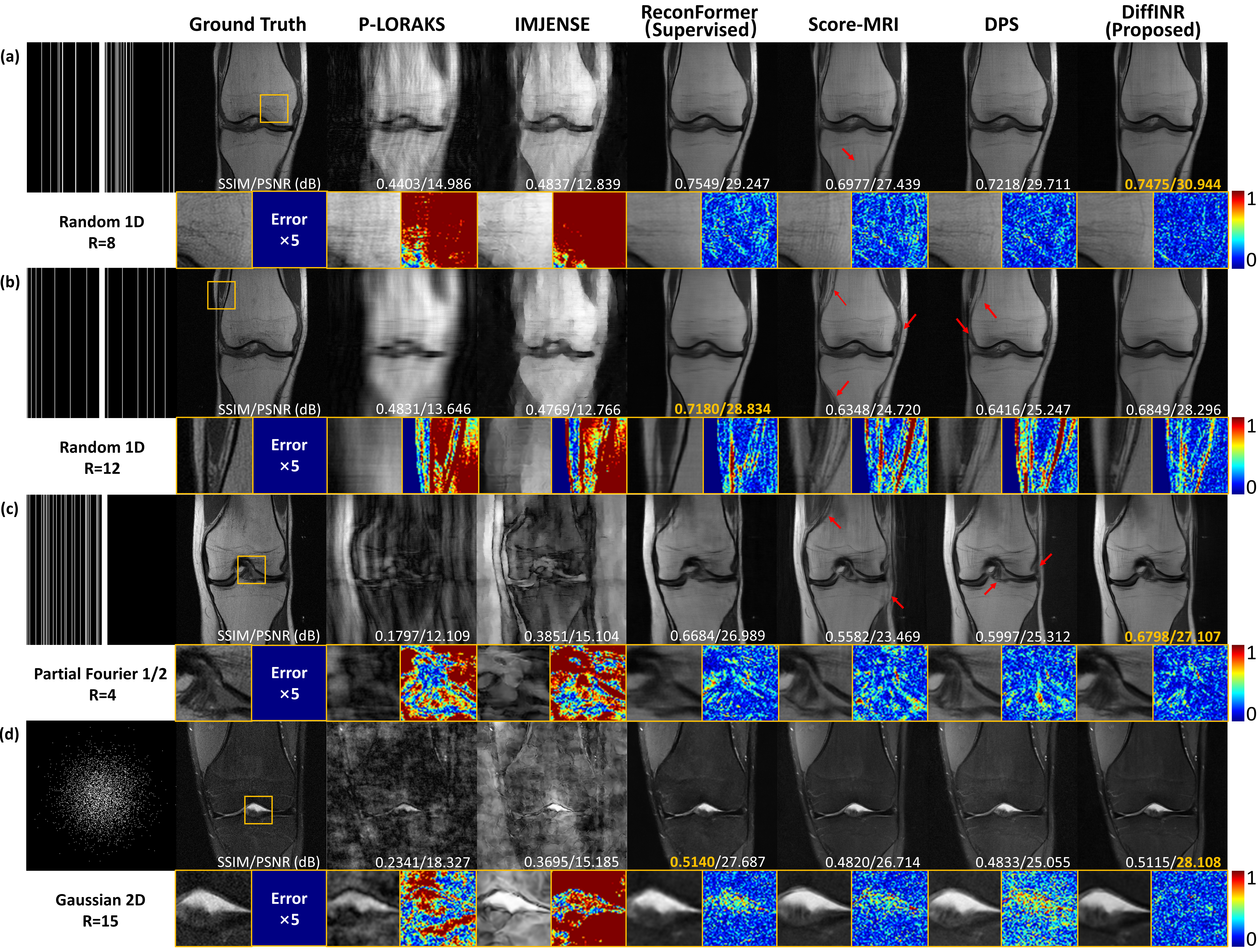}
  \caption{Comparisons of different methods on the single-channel knee dataset. (a) and (b) are reconstruction results of Random 1D sampling with R=8 (ACS=12) and R=12 (ACS=12), respectively. (c) are the results of 1/2 Partial Fourier under-sampling. (d) Gaussian 2D results at R=15. Red arrows point to the artifacts on the magnitude images reconstructed by Score-MRI and DPS. SSIM and PSNR are reported and the best results are in bold and yellow. The ×5 error maps are displayed.}
  \label{fig:XXX}
\end{figure*}

\begin{figure*}[t!]
  \centering
  \includegraphics[width=1\textwidth]{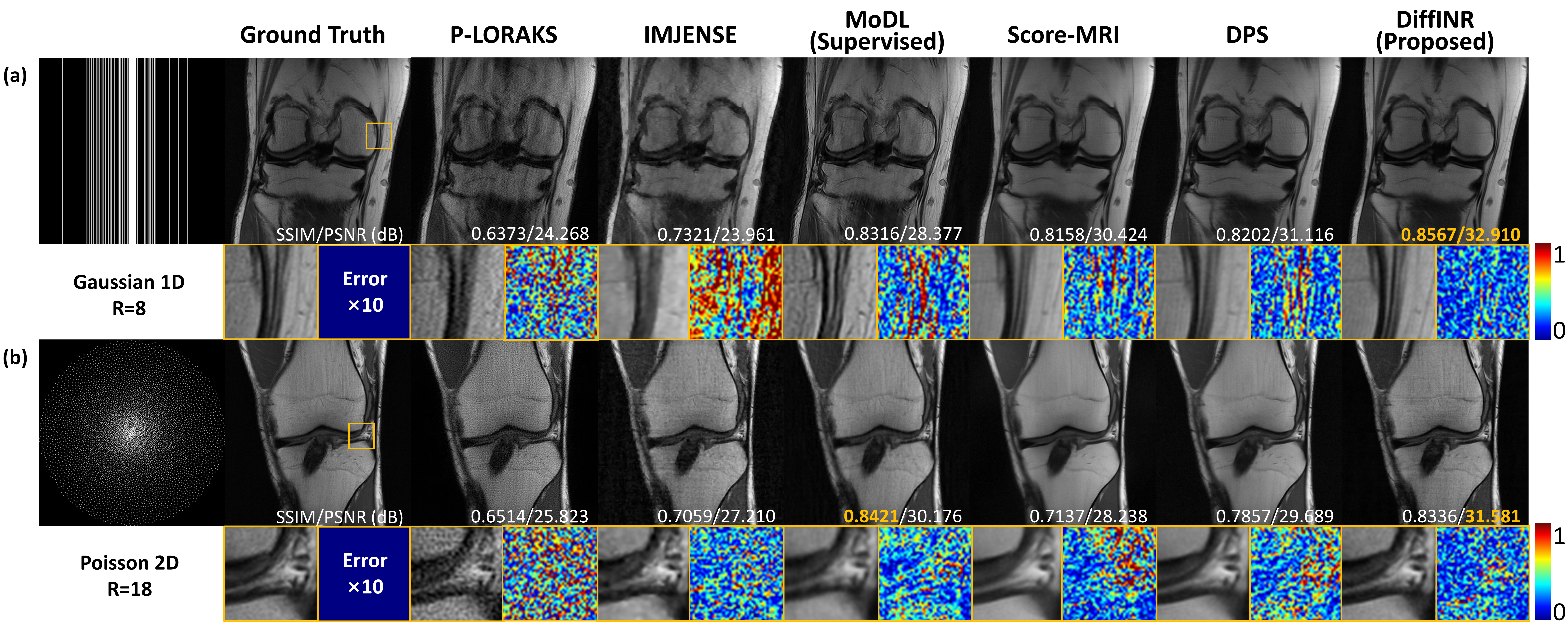}
  \caption{Comparsion results of different methods on the multi-channel dataset with different acceleration factors and sampling masks. (a) is the reconstruction result under Gaussian 1D sampling masks with R=8 (ACS=12). (b) is the result of Poisson 2D with R=18. Quantitative evaluation metrics (SSIM and PSNR) are reported and the best results are emphasized in bold and yellow. The ×10 error maps are displayed.}
  \label{fig:XXX}
\end{figure*}

\subsection{INR Guided Posterior Sampling}
To solve the DC problem in (12), currently there are two strategies, which are direct DC projection and diffusion posterior sampling (DPS).
Score-MRI is a representative DC projection method that tries to directly project the noise image $x_{t-1}$ to the measurement space.
However, there is a relatively large error in direct projection methods because of the inaccurate forward acquisition model between $x_{t-1}$ and $y$, which makes projection-based methods easily getting to the local minimum and sensitive to the noise perturbations.
Additionally, DPS has a high uncertainty due to insufficient DC constraint, which is also called the soft DC operation \cite{ref24}.
Thus, the limitations of current DC operators might lead to an unstable and biased estimation.

In this study, we propose to use INR to guide the DC operation during posterior sampling with a two-stage learning strategy. 
Specifically, INR learns the image $x_{0|t-1}$ to embed the prior information into the network.
Then, we refine $x_{0|t-1}$ based on measurement data to ensure data fidelity.
In this way, we are able to enhance the intermediate $x_{t-1}$ for DC through iterative refinement of the parameters within the INR network, rather than tuning the update step size in DPS.
Leveraging the powerful representation capability of INR \cite{ref21}, we can achieve a highly faithful refinement result $x_{0|t-1}$ for specific scanning protocols.
Detailed explanations of the two-stage INR learning are outlined below.

\textit{1) Stage 1: Prior Embedding}

In Stage 1, we deploy an MLP to map the spatial coordinates to the corresponding signal intensities in the prior image $x_{0|t-1}$.
The MLP is randomly initialized and optimized by the following loss function:
\begin{equation}
  \label{deqn_ex13}
  \phi^{pr}=\underset{\phi}{\mathrm{argmin}}\left\|P_{\phi}(d)-x_{0|t-1}\right\|_{2}^{2},
\end{equation}
where $P_{\phi}$ denotes the MLP network with parameters $\phi$ and $d$ denotes the spatial coordinates of the prior image.
Following the prior embedding, the prior information of $x_{0|t-1}$ is integrated into the MLP network’s weights $\phi^{pr}$, serving as the initial weights for the DC refinement in Stage 2.

\textit{2) Stage 2: Data Consistency Refinement}

In Stage 2, we initiate training with the prior-embedded MLP $P_{\phi^{pr}}$.
Given the prior embedded INR, we are capable of representing $x_{0|t-1}$ as $P_{\phi^{pr}}(d)$ with input coordinates $d$.
During the DC refinement stage, our primary goal is to minimize the difference between $AP_{\phi^{pr}}(d)$ and $y$.
The optimization objective can be formulated as:
\begin{equation}
  \label{deqn_ex14}
  \phi^{*}=\underset{\phi}{\mathrm{argmin}}\left\|AP_{\phi}(d)-y\right\|_{1},
\end{equation}
The network $P_{\phi}$ is trained by an $L_{1}$ loss, initializing the weights with those  embedded from the prior image $\phi^{pr}$.
Finally, the output image can be reconstructed by inferring the spatial coordinates in the trained network, which is $\hat{x}_{0|t-1}=P_{\phi^{*}}(d)$.

Incorporating both prior embedding and DC refinement, $\hat{x}_{0|t-1}$ contains the prior distribution and maintains high fidelity to the data.
To proceed the following posterior sampling steps, $\hat{x}_{0|t-1}$ is remapped back to the diffusion model by adding proper noise:
\begin{equation}
  \label{deqn_ex15}
  x_{t-1}=\sqrt{\bar{\alpha}_{t-1}}\hat{x}_{0|t-1}+\sqrt{1-\bar{\alpha}_{t-1}}\epsilon_{t-1},
\end{equation}
where $\epsilon_{t-1}\in N(0,I)$ is the noise scale at timestep $t-1$.
\subsection{Implementation Details}
We employ a pre-trained diffusion model as referenced in \cite{ref27} to estimate the score value for each sampling step without retraining.
Similar to the previous work in \cite{ref24}, we introduce the INR-based DC operation via a skipped-step mechanism at the start timestep $t^{*}$ with an interval of $k$.
In our experiments, the diffusion sampling process comprises a total number of $T = 2000$, with the INR commencing at timestep $t^{*} = 1200$ and an interval $k = 50$.
Our INR architecture consists of a tiny MLP that includes one input layer, two hidden layers, and one output layer.
Each of the hidden layers has 64 neurons, and all layers, except for the final one, utilize the ReLU activation function. 
Functionally, we deploy two separate MLPs to represent the real and imaginary components of the MRI images.
We incorporate hash encoding to effectively represent the features of the input coordinates.
The hyperparameters of our hash encoding are following the default parameters in \cite{ref14}.
In the prior embedding stage, the number of iterations is $250$ and the learning rate is $1×10^{-3}$.
In the DC refinement stage, the number of iterations is 250 and the learning rate is $1×10^{-5}$.
All the hyperparameters are trained with the Adam optimizer \cite{ref53}.

All the experiments are implemented using PyTorch 1.11.0 in Python 3.7 on a workstation with an Intel i7-12700F processor, 32GB RAM, and an NVIDIA GeForce RTX 3090GPU with 24 GB memory.

\section{Experiments and Results}
\subsection{Datasets}
Our experiments were conducted using the fastMRI knee dataset \cite{ref54,ref55}.
We obtained all the fastMRI datasets from the NYU fastMRI Initiative database with approval from the New York University School of Medicine Institutional Review Board.
The validation knee dataset contains 199 proton density (PD)/proton density fat suppression (PDFS) volumes.
Specifically, the knee data was acquired using a 2D turbo spin-echo sequence with a matrix size of $320\times320$.
For single-channel reconstruction, we randomly chose 40 slices from 15 volumes.
For multi-channel (15 channels) coil reconstruction, we randomly chose 20 slices from 8 volumes.

\subsection{Performance Evaluation}
DiffINR was compared with the following previous algorithms: (1) scan-specific methods P-LORAKS \cite{ref56,ref57}, and IMJENSE \cite{ref21}, (2) supervised DL methods MoDL \cite{ref2} and ReconFormer \cite{ref58}, and (3) diffusion-based methods Score-MRI \cite{ref27}, and DPS \cite{ref30}.
Specifically, P-LORAKS is a conventional calibration-less method. 
IMJENSE is a scan-specific INR-based method.
ReconFormer is an end-to-end DL method for single-channel MRI data reconstruction.
MoDL is a supervised model-based method for multi-channel reconstruction.
The two supervised DL methods were retrained using the fastMRI single-channel and multi-channel datasets, respectively.
The diffusion-based methods compared in our study are all posterior sampling methods.
For a fair comparison, we used the pre-trained diffusion model provided by Chung \textit{et al.} \cite{ref27} to estimate the scores for all the posterior sampling methods and used pre-estimated sensitivity maps via ESPIRiT \cite{ref59} for all the multi-channel reconstruction tasks.
We computed peak signal-to-noise ratio (PSNR) and structural similarity index measure (SSIM) for quantitative evaluation.

\subsection{Comparison Results of Single-channel Datasets}

Fig. 2 displays the reconstructed images of sub-sampled single-channel data. 
P-LORAKS and IMJENSE both fail when reconstructing images under such high acceleration rates (R=8 and R=12).
All diffusion-based methods are capable of reconstructing images with relatively less artifacts, which underscores the powerful generative capabilities of diffusion models.
However, as the acceleration factor increases, Score-MRI and DPS start to generate discrepancies with the ground truths, as highlighted by red arrows in Fig. 2.
For example, in Fig. 2 (b), Score-MRI generates some unreal structures and DPS displays a noticable discontinuity artifact that is absent in the fully sampled data.
Visually, DiffINR generates more precise structural details among diffusion models and achieves the highest PSNR and SSIM.
While the ReconFormer performs well across all sampling patterns, it should be noted that it requires training a distinct model for each specific sampling pattern with paired data to maintain such high-quality reconstructions.
In contrast to the supervised ReconFormer, our method delivers superior or comparable performance without the need for pre-training on any under-sampling conditions.
Furthermore, we conducted experiments on various k-space sampling masks, as shown in Fig. 2 (c) and (d), to verify the superiority of our methods under different sampling patterns.
As expected, DiffINR yielded the lowest errors, the highest PSNR and SSIM across all experiments, with a PSNR that was approximately 2dB higher than current diffusion-based methods.

\subsection{Comparison Results of Multi-channel Datasets}

We also compared the methods on multi-channel datasets to assess the general application performance.
Fig. 3 displays the results under (a) 1D and (b) 2D under-sampling conditions.
P-LORAKS and IMJENSE exhibit severe aliasing artifacts under 1D sampling, as shown in Fig. 3 (a).
Similarly, MoDL also fails to effectively remove the artifacts at 1D high under-sampling pattern.
In contrast, diffusion-based methods are successful in reconstructing artifact-free images due to their strong generative capabilities.
When compared to Score-MRI and DPS, our method yields more accurate high-frequency details and better reconstruction quality, as illustrated by the zoomed-in images.
In Fig. 3 (b), the result of our method shows a performance level that is on par with MoDL’s result. However, MoDL’s result visually appears smoother.
It should be noted that at such high acceleration factors, DiffINR demonstrates remarkable robustness by maintaining a PSNR of 30 dB or higher.

\section{Conclusion}
In this study, we proposed DiffINR, an INR-based diffusion posterior sampler for highly accelerated MRI reconstruction.
We introduced a pre-trained diffusion model to generate noiseless prior images, allowing the INR to learn high-quality distribution information.
By incorporating the physical model into the INR, robust DC is realized.
Through iterative INR procedures, the diffusion model produces highly accurate and stable reconstruction results.
Experimental results show that DiffINR outperforms the compared methods.
Additionally, this posterior sampler is training-free and is not limited by the specific diffusion model, facilitating its integration with other pre-trained diffusion models for task-specific reconstruction.


\begin{thebibliography}{99}  
  \bibitem{ref1}M. Lustig, D. Donoho, and J. M. Pauly, ``Sparse MRI: The application of compressed sensing for rapid MR imaging," \textit{Magnetic Resonance in Medicine: An Official Journal of the International Society for Magnetic Resonance in Medicine,} vol. 58, no. 6, pp. 1182-1195, Oct. 2007.  
  \bibitem{ref2}H. K. Aggarwal, M. P. Mani, and M. Jacob, ``MoDL: Model-based deep learning architecture for inverse problems," \textit{IEEE transactions on medical imaging,} vol. 38, no. 2, pp. 394-405, Aug. 2018.  
  \bibitem{ref3}M. Doneva, P. Börnert, H. Eggers, C. Stehning, J. Sénégas, and A. Mertins, ``Compressed sensing reconstruction for magnetic resonance parameter mapping," \textit{Magnetic Resonance in Medicine,} vol. 64, no. 4, pp. 1114-1120, Jun. 2010.
  \bibitem{ref4}H. Gu, B. Yaman, K. Ugurbil, S. Moeller, and M. Akçakaya, ``Compressed Sensing MRI with $\ell$1-Wavelet Reconstruction Revisited Using Modern Data Science Tools," in \textit{2021 43rd Annual International Conference of the IEEE Engineering in Medicine {\&} Biology Society (EMBC),} Nov. 2021, pp. 3596-3600.
  \bibitem{ref5}K. H. Jin, D. Lee, and J. C. Ye, ``A general framework for compressed sensing and parallel MRI using annihilating filter based low-rank Hankel matrix," \textit{IEEE Transactions on Computational Imaging,} vol. 2, no. 4, pp. 480-495, Aug. 2016.
  \bibitem{ref6}S. G. Lingala, Y. Hu, E. DiBella, and M. Jacob, ``Accelerated dynamic MRI exploiting sparsity and low-rank structure: kt SLR," \textit{IEEE transactions on medical imaging,} vol. 30, no. 5, pp. 1042-1054, Jan. 2011.
  \bibitem{ref7}T. Zhang, J. M. Pauly, and I. R. Levesque, ``Accelerating parameter mapping with a locally low rank constraint," \textit{Magnetic resonance in medicine,} vol. 73, no. 2, pp. 655-661, Feb. 2015.
  \bibitem{ref8}B. Zhu, J. Z. Liu, S. F. Cauley, B. R. Rosen, and M. S. Rosen, ``Image reconstruction by domain-transform manifold learning," \textit{Nature,} vol. 555, no. 7697, pp. 487-492, Mar. 2018.
  \bibitem{ref9}K. Hammernik \textit{et al.}, ``Learning a variational network for reconstruction of accelerated MRI data," \textit{Magnetic Resonance in Medicine,} vol. 79, no. 6, pp. 3055-3071, Nov. 2018.
  \bibitem{ref10}K. H. Jin, M. T. McCann, E. Froustey, and M. Unser, ``Deep convolutional neural network for inverse problems in imaging," \textit{IEEE transactions on image processing,} vol. 26, no. 9, pp. 4509-4522, Jun. 2017.
  \bibitem{ref11}G. Ongie, A. Jalal, C. A. Metzler, R. G. Baraniuk, A. G. Dimakis, and R. Willett, ``Deep learning techniques for inverse problems in imaging," \textit{IEEE Journal on Selected Areas in Information Theory,} vol. 1, no. 1, pp. 39-56, May. 2020.
  \bibitem{ref12}D. Ulyanov, A. Vedaldi, and V. Lempitsky, ``Deep image prior," in \textit{Proceedings of the IEEE conference on computer vision and pattern recognition,} 2018, pp. 9446-9454.
  \bibitem{ref13}N. Rahaman \textit{et al.}, ``On the spectral bias of neural networks," in \textit{International conference on machine learning,} vol. 97, Jun. 2019, pp. 5301-5310.
  \bibitem{ref14}Z. Shi, P. Mettes, S. Maji, and C. G. Snoek, ``On measuring and controlling the spectral bias of the deep image prior," \textit{International Journal of Computer Vision,} vol. 130, no. 4, pp. 885-908, Feb. 2022.
  \bibitem{ref15}G. Yüce, G. Ortiz-Jiménez, B. Besbinar, and P. Frossard, ``A structured dictionary perspective on implicit neural representations," in \textit{Proceedings of the IEEE/CVF Conference on Computer Vision and Pattern Recognition,} Jun. 2022, pp. 19228-19238.
  \bibitem{ref16}M. Tancik \textit{et al.}, ``Fourier features let networks learn high frequency functions in low dimensional domains," \textit{Advances in Neural Information Processing Systems,} vol. 33, pp. 7537-7547, 2020.
  \bibitem{ref17}B. Mildenhall, P. P. Srinivasan, M. Tancik, J. T. Barron, R. Ramamoorthi, and R. Ng, ``Nerf: Representing scenes as neural radiance fields for view synthesis," \textit{Communications of the ACM,} vol. 65, no. 1, pp. 99-106, Jan. 2021.
  \bibitem{ref18}J. Wang, Y. Wang, J. Deng, and D. Liu, ``Unsupervised Coordinate-Based Neural Network for Electrical Impedance Tomography," \textit{IEEE Transactions on Computational Imaging,} vol. 9, pp. 1213-1225, Dec. 2023.
  \bibitem{ref19}T. Müller, A. Evans, C. Schied, and A. Keller, ``Instant neural graphics primitives with a multiresolution hash encoding," \textit{ACM Transactions on Graphics (ToG),} vol. 41, no. 4, pp. 1-15, Jul. 2022.
  \bibitem{ref20}L. Shen, J. Pauly, and L. Xing, ``NeRP: implicit neural representation learning with prior embedding for sparsely sampled image reconstruction," \textit{IEEE Transactions on Neural Networks and Learning Systems,} vol. 35, no. 1, pp. 770-782, Jan. 2024.
  \bibitem{ref21}R. Feng \textit{et al.}, ``IMJENSE: scan-specific implicit representation for joint coil sensitivity and image estimation in parallel MRI," \textit{IEEE Transactions on Medical Imaging,} vol. 43, no. 4, pp. 1539-1553, Apr. 2024.
  \bibitem{ref22}D. Strong and T. Chan, ``Edge-preserving and scale-dependent properties of total variation regularization," \textit{Inverse problems,} vol. 19, no. 6, p. S165, Nov. 2003.
  \bibitem{ref23}H. Chung, B. Sim, and J. C. Ye, ``Come-closer-diffuse-faster: Accelerating conditional diffusion models for inverse problems through stochastic contraction," in \textit{Proceedings of the IEEE/CVF Conference on Computer Vision and Pattern Recognition,} Jun. 2022, pp. 12413-12422.
  \bibitem{ref24}B. Song, S. M. Kwon, Z. Zhang, X. Hu, Q. Qu, and L. Shen, ``Solving inverse problems with latent diffusion models via hard data consistency," in \textit{The Twelfth International Conference on Learning Representations,} 2024.
  \bibitem{ref25}Y. Song, L. Shen, L. Xing, and S. Ermon, ``Solving inverse problems in medical imaging with score-based generative models," \textit{arXiv preprint arXiv:2111.08005,} 2021.
  \bibitem{ref26}C. Peng, P. Guo, S. K. Zhou, V. M. Patel, and R. Chellappa, ``Towards performant and reliable undersampled MR reconstruction via diffusion model sampling," in \textit{International Conference on Medical Image Computing and Computer-Assisted Intervention,} Sep. 2022, pp. 623-633.
  \bibitem{ref27}H. Chung and J. C. Ye, ``Score-based diffusion models for accelerated MRI," Medical image analysis, vol. 80, p. 102479, Aug. 2022.
  \bibitem{ref28}A. Jalal, M. Arvinte, G. Daras, E. Price, A. G. Dimakis, and J. Tamir, ``Robust compressed sensing mri with deep generative priors," \textit{Advances in Neural Information Processing Systems,} vol. 34, pp. 14938-14954, 2021.
  \bibitem{ref29}H. Chung, B. Sim, D. Ryu, and J. C. Ye, ``Improving diffusion models for inverse problems using manifold constraints," \textit{Advances in Neural Information Processing Systems,} vol. 35, pp. 25683-25696, 2022.
  \bibitem{ref30}H. Chung, J. Kim, M. T. Mccann, M. L. Klasky, and J. C. Ye, ``Diffusion posterior sampling for general noisy inverse problems," in \textit{The Eleventh International Conference on Learning Representations,} 2023.
  \bibitem{ref31}H. Chung, S. Lee, and J. C. Ye, ``Decomposed Diffusion Sampler for Accelerating Large-Scale Inverse Problems," in \textit{The Twelfth International Conference on Learning Representations,} Vienna, Austria, 2024.
  \bibitem{ref32}G. Luo, M. Blumenthal, M. Heide, and M. Uecker, ``Bayesian MRI reconstruction with joint uncertainty estimation using diffusion models," \textit{Magnetic Resonance in Medicine,} vol. 90, no. 1, pp. 295-311, Mar. 2023.
  \bibitem{ref33}Y. Korkmaz, T. Cukur, and V. M. Patel, ``Self-supervised MRI Reconstruction with Unrolled Diffusion Models," \textit{Medical Image Computing and Computer Assisted Intervention,} Cham, Germany, 2023, pp. 491-501.
  \bibitem{ref34}Y. Xie and Q. Li, ``Measurement-Conditioned Denoising Diffusion Probabilistic Model for Under-Sampled Medical Image Reconstruction," \textit{Medical Image Computing and Computer Assisted Intervention,} Cham, Germany, 2022, pp. 655-664.
  \bibitem{ref35}Y. Cao, L. Wang, J. Zhang, H. Xia, F. Yang, and Y. Zhu, ``Accelerating multi-echo MRI in k-space with complex-valued diffusion probabilistic model," in \textit{2022 16th IEEE International Conference on Signal Processing (ICSP),} 2022, vol. 1, pp. 479-484.
  \bibitem{ref36}A. Güngör \textit{et al.}, ``Adaptive diffusion priors for accelerated MRI reconstruction," \textit{Medical Image Analysis,} vol. 88, pp. 102872, Aug. 2023.
  \bibitem{ref37}C. Cao \textit{et al.}, ``High-Frequency Space Diffusion Model for Accelerated MRI," \textit{IEEE Transactions on Medical Imaging,} pp. 1-1, Jan. 2024.
  \bibitem{ref38}Z.-X. Cui \textit{et al.}, ``Self-score: Self-supervised learning on score-based models for mri reconstruction," \textit{arXiv preprint arXiv:2209.00835,} 2022.
  \bibitem{ref39}C. Cao \textit{et al.}, ``SPIRiT-diffusion: SPIRiT-driven score-based generative modeling for vessel wall imaging," \textit{arXiv preprint arXiv:2212.11274,} 2022.
  \bibitem{ref40}M. Lustig and J. M. Pauly, ``SPIRiT: Iterative self-consistent parallel imaging reconstruction from arbitrary k-space," \textit{Magnetic Resonance in Medicine,}  vol. 64, no. 2, pp. 457-471, 2010..
  \bibitem{ref41}Z. Tu \textit{et al.}, ``WKGM: weighted k-space generative model for parallel imaging reconstruction," \textit{NMR in Biomedicine,} vol. 36, no. 11, p. e5005, Aug. 2023.
  \bibitem{ref42}H. Peng \textit{et al.}, ``One-Shot Generative Prior in Hankel-k-Space for Parallel Imaging Reconstruction," \textit{IEEE Transactions on Medical Imaging,} vol. 42, no. 11, pp. 3420-3435, Nov. 2023.
  \bibitem{ref43}R. Liu, Y. Sun, J. Zhu, L. Tian, and U. S. Kamilov, ``Recovery of continuous 3d refractive index maps from discrete intensity-only measurements using neural fields," \textit{Nature Machine Intelligence,} vol. 4, no. 9, pp. 781-791, Sep. 2022.
  \bibitem{ref44}J. J. Park, P. Florence, J. Straub, R. Newcombe, and S. Lovegrove, ``Deepsdf: Learning continuous signed distance functions for shape representation," in \textit{Proceedings of the IEEE/CVF conference on computer vision and pattern recognition,} Jun. 2019, pp. 165-174.
  \bibitem{ref45}V. Saragadam, D. LeJeune, J. Tan, G. Balakrishnan, A. Veeraraghavan, and R. G. Baraniuk, ``Wire: Wavelet implicit neural representations," in \textit{Proceedings of the IEEE/CVF Conference on Computer Vision and Pattern Recognition,} 2023, pp. 18507-18516.
  \bibitem{ref46}Y. Lu, Z. Wang, M. Liu, H. Wang, and L. Wang, ``Learning spatial-temporal implicit neural representations for event-guided video super-resolution," in \textit{Proceedings of the IEEE/CVF Conference on Computer Vision and Pattern Recognition,} Jun. 2023, pp. 1557-1567.
  \bibitem{ref47}Q. Wu, R. Feng, H. Wei, J. Yu, and Y. Zhang, ``Self-supervised coordinate projection network for sparse-view computed tomography," \textit{IEEE Transactions on Computational Imaging,} vol. 9, pp. 517-529, Jun. 2023.
  \bibitem{ref48}A. W. Reed \textit{et al.}, ``Dynamic ct reconstruction from limited views with implicit neural representations and parametric motion fields," in \textit{Proceedings of the IEEE/CVF International Conference on Computer Vision,} Oct. 2021, pp. 2258-2268.
  \bibitem{ref49}J. Xu \textit{et al.}, ``Nesvor: Implicit neural representation for slice-to-volume reconstruction in mri," \textit{IEEE Transactions on Medical Imaging,} vol. 42, no. 6, pp. 1707-1719, Jun. 2023.
  \bibitem{ref50}B. Efron, ``Tweedie’s formula and selection bias," \textit{Journal of the American Statistical Association,} vol. 106, no. 496, p. 1602, 2011.
  \bibitem{ref51}K. Kim, T. Kwon, and J. C. Ye, ``Noise distribution adaptive self-supervised image denoising using tweedie distribution and score matching," in \textit{Proceedings of the IEEE/CVF Conference on Computer Vision and Pattern Recognition,} Jun. 2022, pp. 2008-2016.
  \bibitem{ref52}R. Courant, ``Variational methods for the solution of problems of equilibrium and vibrations," in \textit{Bull. Amer. Math. Soc.,} vol. 49, no. 12, pp. 1-23, 1943.
  \bibitem{ref53}D. P. Kingma and J. Ba, ``Adam: A method for stochastic optimization," in \textit{arXiv preprint arXiv:1412.6980,} 2014.
  \bibitem{ref54}F. Knoll \textit{et al.}, ``fastMRI: A publicly available raw k-space and DICOM dataset of knee images for accelerated MR image reconstruction using machine learning," in \textit{Radiology: Artificial Intelligence,} vol. 2, no. 1, p. e190007, 2020.
  \bibitem{ref55}J. Zbontar \textit{et al.}, ``fastMRI: An open dataset and benchmarks for accelerated MRI," \textit{arXiv preprint arXiv:1811.08839,} 2018.
  \bibitem{ref56}J. P. Haldar and J. Zhuo, ``P‐LORAKS: low‐rank modeling of local k‐space neighborhoods with parallel imaging data," \textit{Magnetic resonance in medicine,} vol. 75, no. 4, pp. 1499-1514, May. 2016.
  \bibitem{ref57}T. H. Kim and J. P. Haldar, ``“LORAKS software version 2.0: Faster implementation and enhanced capabilities," \textit{niversity of Southern California, Los Angeles, CA, Tech. Rep. USC-SIPI-443,}2018.
  \bibitem{ref58}P. Guo, Y. Mei, J. Zhou, S. Jiang, and V. M. Patel, ``ReconFormer: Accelerated MRI reconstruction using recurrent transformer," \textit{IEEE transactions on medical imaging,} vol. 43, no. 1, pp. 582-593, Jan. 2024.
  \bibitem{ref59}M. Uecker \textit{et al.}, ``ESPIRiT—an eigenvalue approach to autocalibrating parallel MRI: where SENSE meets GRAPPA," \textit{Magnetic resonance in medicine,} vol. 71, no. 3, pp. 990-1001, May. 2014.
  \end{thebibliography}
\end{document}